# Testing the running of the QCD coupling in particle energy spectra

Sergio Lupia[1] and Wolfgang Ochs[2]

*Max-Planck-Institut für Physik, Werner-Heisenberg-Institut*
*Föhringer Ring 6, D–80805 München, Germany*

## Abstract

The spectra of charged particles in the energy variable $\xi = \log 1/x$ ("the hump backed plateau") are analyzed in terms of moments in order to investigate the sensitivity to the running of $\alpha_s$ in particular towards small scales, and to further test thereby the local parton hadron duality picture. The modified leading log approximation (MLLA) of QCD provides a very satisfying description of the $e^+e^-$ data in the full range of c.m. energies from about 1 up to 91 GeV. The same calculation but with fixed coupling $\alpha_s$ yields moments which are inconsistent with the data already close to threshold. In particular the soft particles ($E \leq 1$ GeV) reflect directly the running of $\alpha_s$.

---

[1] E-mail address: lupia@mppmu.mpg.de
[2] E-mail address: wwo@dmumpiwh

## 1. Introduction

The predictions from perturbative QCD on the intrinsic structure of jets based on the idea of a soft confinement mechanism [1, 2] have been remarkably successful. In particular, the prediction of the "hump backed plateau", i.e., an approximately Gaussian distribution in $\log(P_{\text{jet}}/p_{\text{hadron}})$ with a suppression of low momentum particles from the soft gluon interference [3, 4, 5, 6] has been confirmed experimentally. This demonstrated the close similarity of parton and hadron spectra, provided the parton cascade is evolved down to a cutoff scale of the order of the hadronic masses $Q_0 \simeq m_h$. In this way, an almost quantitative description of the momentum spectra in the PETRA/PEP energy range (see, e.g., [7]) and at LEP[8, 9, 10, 11] has been obtained ("Local Parton Hadron Duality"[12]). The first data from HERA[13, 14] confirm this picture as well.

In this paper we investigate the sensitivity of the energy spectrum of final state particles to the running of the coupling constant $\alpha_s$. The effect of the running $\alpha_s$ is strongest for soft particles and at low energies and its verification can therefore provide an additional hint on the applicability of perturbative QCD towards small scales. To this end, we use a consistent kinematical scheme to relate partons and hadrons. We analyze the spectra in terms of moments for which explicit analytical expressions can be derived for fixed and running coupling and we extend the analysis down to low energy $\sqrt{s}$ of order 1 GeV.

An appropriate theoretical framework is the evolution equation for parton densities in the modified leading log approximation (MLLA) (for a recent review see [2]). It determines the parton densities and moments in absolute magnitude by the boundary condition at threshold in terms of two parameters, the QCD scale $\Lambda$ and the $p_t$ cutoff $Q_0$. This is different from deep inelastic scattering, where an unknown parton distribution has to be provided at a particular scale. The asymptotic predictions at high energies agree with those from the double log approximation (DLA) which neglects the energy-momentum constraints[3, 4]. The Next-to-Leading corrections of relative order $\sqrt{\alpha_s}$ for moments are found large[5]. The MLLA results[6] sum all orders of $\sqrt{\alpha_s}$ which allows the extrapolation to threshold but not all contributions beyond Next-to-Leading order are taken into account. More details of this study will be published elsewhere, see also [15].

## 2. The theoretical framework

*MLLA Evolution Equations.* The evolution equation of single parton inclusive energy distribution in the framework of MLLA is given by:

$$\frac{d}{d\log\Theta} x \bar{D}_A^B(x, \log P\Theta) = \sum_{C=q,\bar{q},g} \int_0^1 dz \frac{\alpha_s(k_t)}{2\pi} \Phi_A^C(x) \left[\frac{x}{z} \bar{D}_C^B(\frac{x}{z}, \log(zP\Theta))\right] \quad (1)$$

with the boundary condition at threshold $P\Theta = Q_0$:

$$x \bar{D}_A^B(x, \log Q_0) = \delta(1-x)\delta_A^B \quad (2)$$

Here $P$ and $\Theta$ denote the primary parton energy and jet opening angle respectively, $x$ is the energy fraction carried by the produced parton, $\Phi_i^j(z)$ are the parton splitting kernels and $i,j$, $A$, $B$, $C$ label quarks, antiquarks and gluons. The QCD running coupling is given by its one-loop expression $\alpha_s(k_t) = 2\pi/b\log(k_t/\Lambda)$ with $b \equiv (11N_c - 2n_f)/3$, $\Lambda$ the QCD-scale and $N_c$ and $n_f$ the number of colors and of flavors respectively. The scale of the coupling is given by the transverse momentum $k_t \simeq z(1-z)P\Theta$. The shower evolution is cut off by $Q_0$, such that $k_t \geq Q_0$.

The integral equation can be solved by Mellin transform:

$$D_\omega(Y,\lambda) = \int_0^1 \frac{dx}{x} x^\omega [x\bar{D}(x,Y)] = \int_0^Y d\xi e^{-\xi\omega} D(\xi,Y,\lambda) \qquad (3)$$

with $Y = \log\frac{P\Theta}{Q_0} \simeq \log\frac{P}{Q_0}$, $\lambda = \log\frac{Q_0}{\Lambda}$, $\xi = \log\frac{P}{k}$ and parton energy $k$.

In flavor space the valence quark and $(\pm)$ mixtures of sea quarks and gluons evolve independently with different "eigenfrequencies". At high energies, the dominant contribution to the inclusive spectrum comes from the "plus"-term, which we denote by $D_\omega(Y,\lambda) \equiv D_\omega^+(Y,\lambda)$. In an approximation where only the leading singularity in $\omega$-space plus a constant term is kept, it satisfies the following evolution equation:

$$\left(\omega + \frac{d}{dY}\right)\frac{d}{dY}D_\omega(Y,\lambda) - 4N_c\frac{\alpha_s(Y+\lambda)}{2\pi}D_\omega(Y,\lambda) \qquad (4)$$
$$= -a\left(\omega + \frac{d}{dY}\right)\frac{\alpha_s(Y+\lambda)}{2\pi}D_\omega(Y,\lambda)$$

where $a = 11N_c/3 + 2n_f/3N_c^2$. Taking $a = 0$ and dropping the r.h.s would yield the DLA evolution equation. The analysis of this paper is based on Eq. (4).

By defining now the anomalous dimension $\gamma_\omega$ according to:

$$D_\omega(Y,\lambda) = D_\omega(Y_0,\lambda)\exp\left(\int_{Y_0}^Y dy\gamma_\omega[\alpha_s(y+\lambda)]\right) \qquad (5)$$

the evolution equation for the inclusive spectrum can be written as a differential equation for $\gamma_\omega$ as follows:

$$(\omega + \gamma_\omega)\gamma_\omega - \frac{4N_c\alpha_s}{2\pi} = -\beta(\alpha_s)\frac{d}{d\alpha_s}\gamma_\omega - 2\eta(\omega+\gamma_\omega) \qquad (6)$$

where $\beta(\alpha_s) = \frac{d}{dY}\alpha_s(Y) \simeq -b\frac{\alpha_s^2}{2\pi}$ and $\eta = a\gamma_0^2/8N_c = a\alpha_s/4\pi$. In general this nonlinear equation has two roots and the solution consists of a superposition of two terms of the type (5).

*Moment analysis for running* $\alpha_s$. Explicit expressions with parameters $Q_0$ and $\Lambda$ can be derived for the multiplicity $\bar{\mathcal{N}}$ and the normalized moments $<\xi^q(Y,\lambda)> = \int d\xi\xi^q D(\xi,Y,\lambda)/\bar{\mathcal{N}}$ or the cumulant moments $\kappa_q(Y,\lambda)$[5, 6]; they are related by $\kappa_1 = <\xi> = \bar{\xi}$, $\kappa_2 \equiv \sigma^2 = <(\xi-\bar{\xi})^2>$, $\kappa_3 = <(\xi-\bar{\xi})^3>$, $\kappa_4 = <(\xi-\bar{\xi})^4> -3\sigma^4$, ...; also one introduces the reduced cumulants $k_q \equiv \kappa_q/\sigma^q$, in particular the skewness $s = k_3$ and the kurtosis $k = k_4$. The cumulants $\kappa_q$ can be obtained from:

$$\kappa_q(Y,\lambda) = \left(-\frac{\partial}{\partial\omega}\right)^q \log D_\omega(Y,\lambda)\bigg|_{\omega=0}, \qquad (7)$$

also $\bar{\mathcal{N}}_E = D_\omega|_{\omega=0}$. At high energies, one term of the form (5) dominates and one obtains

$$\kappa_q(Y) = \kappa_q(Y_0) + \int_{Y_0}^Y dy\left(-\frac{\partial}{\partial\omega}\right)^q \gamma_\omega[\alpha_s(y)]\bigg|_{\omega=0} \qquad (8)$$

This equation shows the direct dependence of the cumulants on $\alpha_s(Y)$. For fixed $\alpha_s$, for example, one obtains directly $\kappa_q(Y) \propto Y$ for high energies. In our applications, we take into account the boundary conditions at threshold $Y_0 = 0$, eq. (2), which yields $\kappa_q(Y_0 = 0) = 0$.

The eqs. (5) and (8) are well suited to discuss the dependence on the number of flavors $n_f$. The simplest procedure would consist in taking $\gamma_\omega(y)$ under the integral as step function with $n_f$ as the number of open flavors at energy $y$. In leading order of $\sqrt{Y}$ (, i.e., in the DLA), $n_f$ enters only through the running coupling $\alpha_s(y, n_f)$, at non leading level also through the parameter $a$ (from the splitting function $\Phi_A^C$). Restricting our discussion to the leading term we note that in the original eq. (1) the scale of $\alpha_s$ is $k_t \simeq z(1-z)P < P/4$ and the maximal scale $P$ appearing in eqs. (5) and (8) originates from a kinematical approximation. In our comparison with data we choose either $n_f = 3$ throughout as in previous analyses or we include the effect of heavy quarks in eq. (5) and (8) with $n_f$ changing for momenta $P_Q = 4m_Q$. A more accurate treatment of scales would have to take into account two-loop contributions besides the above kinematical effects and possibly explicit mass dependence of the coupling[16].

The evolution equation for $D_\omega(Y, \lambda)$ can be analytically solved and one obtains[2] a linear superposition of two hypergeometric distributions. Moments of order $q$ can then be obtained through the formula $<\xi^q> = [(-\frac{\partial}{\partial \omega})^q D_\omega(Y, \lambda)]|_{\omega=0}/\mathcal{N}$ and can be written in general as:

$$<\xi^q> = \frac{1}{\mathcal{N}} \sum_{k=0}^{q} \binom{q}{k} (N_1 L_k^{(q)} + N_2 R_k^{(q)}) \qquad (9)$$

where $N_1$, $N_2$, $L_k^{(q)}$ and $R_k^{(q)}$ are known functions of $a$, $b$, $Y + \lambda$ and $\lambda$. So they depend on the two parameters $Q_0$ and $\Lambda$[6]. We found, as discussed below, the best agreement for the "limiting spectrum", where the two parameters coincide, $Q_0 = \Lambda$, i.e., $\lambda = 0$. In this case, the formulae simplify and the moments can be expressed[6] in terms of the parameter $B \equiv a/b$ and the variable $z \equiv \sqrt{16 N_c Y/b}$ as:

$$\frac{<\xi^q>}{Y^q} = P_0^{(q)}(B+1, B+2, z) + \frac{2}{z} \frac{I_{B+2}(z)}{I_{B+1}(z)} P_1^{(q)}(B+1, B+2, z) \qquad (10)$$

where $P_0^{(q)}$ and $P_1^{(q)}$ are polynomials of order $2(q-1)$ in $z$. These moments have an expansion $<\xi^q>/Y^q = c_0^{(q)} + c_1^{(q)}/\sqrt{Y} + \ldots$ [3]. The higher orders of this series are still numerically sizeable at LEP energies (10% contribution from NNLO to $\bar{\xi}$ and $\sigma^2$) and increase towards lower energies. We therefore included the full expression, also for $P_i^{(q)}$; explicit results for the moments for $q \leq 3$ can be found in [6] and for $q = 4$ in [15]. The average multiplicity of partons (zeroth-order moment of the inclusive momentum spectrum) for the Limiting Spectrum is given by:

$$\bar{\mathcal{N}}_{LS} = \Gamma(B) \left(\frac{z}{2}\right)^{1-B} I_{B+1}(z) \,. \qquad (11)$$

This expression does not fulfill the boundary condition $\bar{\mathcal{N}}(0) = 1$. Therefore in our fits we added a constant term[4].

---

[3]We checked that $\bar{\xi}$ and $\sigma^2$ agree in NLO and $s$ in LO with the result obtained by Fong and Webber[5], i.e., agreement of $<\xi^q>$ in NLO.

[4]The general solution considered in [6] yields $dD_\omega/dY|_{Y=0} = -B/\lambda$; therefore the multiplicity for

*Moment analysis for fixed $\alpha_s$.* In order to study the sensitivity of the inclusive energy distribution to the running of the coupling, let us build a new model based again upon the MLLA but with the coupling $\alpha_s$ or, equivalently, the anomalous dimension $\gamma_0 = \sqrt{6\alpha_s/\pi}$, kept fixed as a free parameter replacing $\Lambda$. In this case, the evolution equation for the Mellin transform, $D_\omega(Y,\lambda)$, eq. (4), becomes:

$$\left(\omega + \frac{d}{dY}\right)\frac{d}{dY}D_\omega(Y,\lambda) - \gamma_0^2 D_\omega(Y,\lambda) = -\eta\left(\omega + \frac{d}{dY}\right)D_\omega(Y,\lambda). \tag{12}$$

One can then perform explicitly the inverse Mellin transform

$$\frac{\partial^2 D(\xi,Y,\lambda)}{\partial\xi\partial Y} + \frac{\partial^2 D(\xi,Y,\lambda)}{dY^2} - \gamma_0^2 D(\xi,Y,\lambda) = -2\eta\frac{\partial D(\xi,Y,\lambda)}{\partial\xi} - 2\eta\frac{\partial D(\xi,Y,\lambda)}{\partial Y} \tag{13}$$

Through the simple transformation: $D(\xi,Y,\lambda) = \tilde{D}(\xi,Y,\lambda)\exp[-2\eta(Y-\xi)]$, one gets rid of the linear terms and $\tilde{D}(\xi,Y,\lambda)$ obeys eq. (13) with $\eta = 0$, i.e., the DLA equation. Using its known solution[2], we find for the inclusive energy spectrum in MLLA with fixed-$\alpha_s$

$$D(\xi,Y,\lambda) = \gamma_0\sqrt{\frac{Y-\xi}{\xi}}I_1\left(2\gamma_0\sqrt{\xi(Y-\xi)}\right)e^{-2\eta(Y-\xi)} \tag{14}$$

The exact form of cumulants can also be obtained in a straightforward way from the evolution equation for the anomalous dimension $\gamma_\omega$ given in eq. (6). For fixed $\alpha_s$, the first term on the r.h.s. drops out; we obtain an algebraic equation with two solutions $\gamma_{\omega\pm}$ and we find:

$$D_\omega(Y,\lambda) = \frac{\gamma_{\omega-}}{\gamma_{\omega-}-\gamma_{\omega+}}e^{\gamma_{\omega+}Y} + \frac{\gamma_{\omega+}}{\gamma_{\omega+}-\gamma_{\omega-}}e^{\gamma_{\omega-}Y} \tag{15}$$

$$\gamma_{\omega\pm} = \frac{1}{2}\left[-(\omega+2\eta) \pm \sqrt{(\omega-2\eta)^2 + 4\gamma_0^2}\right] \tag{16}$$

The results for the moments can be obtained easily from eq. (7); for example, we find for the average multiplicity $\bar{\mathcal{N}}_{fix}$:

$$\bar{\mathcal{N}}_{fix} = \left[\cosh(\bar{\gamma}_0 Y) + \frac{\eta}{\bar{\gamma}_0}\sinh(\bar{\gamma}_0 Y)\right]\exp(-\eta Y) \tag{17}$$

with $\bar{\gamma}_0 \equiv \sqrt{\gamma_0^2 + \eta^2}$; it reduces for $\eta = 0$ to the well-known DLA result $\cosh(\gamma_0 Y)$, and for the first moment

$$\bar{\xi}_{fix} = [A\cosh(\bar{\gamma}_0 Y) + B\sinh(\bar{\gamma}_0 Y)]\frac{\exp(-\eta Y)}{\bar{\mathcal{N}}_{fix}} \tag{18}$$

where $A = Y(\bar{\gamma}_0^2 + \eta^2)/2\bar{\gamma}_0^2$ and $B = (2\bar{\gamma}_0^2(\eta Y - 1) - \eta^2)/2\bar{\gamma}_0^3$. By choosing the + term in eq. (16), one selects the leading high energy contribution which also dominates the "limiting spectrum" for the running $\alpha_s$ model. In this limit, according to eq. (8), one gets:

$$\bar{\mathcal{N}}_{fix} = \frac{1}{2}\left(1 + \frac{\eta}{\bar{\gamma}_0}\right)\exp([\bar{\gamma}_0 - \eta]Y) \tag{19}$$

---

general $\lambda$ has an unphysical behavior near threshold $d\bar{\mathcal{N}}(Y)/dY|_{Y=0} < 0$ with $\bar{\mathcal{N}}(0) = 1$. For $\lambda = 0$, $\bar{\mathcal{N}}(0) = 0$ and $d\bar{\mathcal{N}}(Y)/dY|_{Y=0} > 0$.

$$\bar{\xi}_{fix} = \left[1 + \frac{\eta}{\bar{\gamma}_0}\right] \frac{Y}{2} \quad , \quad \sigma^2_{fix} = \frac{\gamma_0^2}{4\bar{\gamma}_0^3} Y \tag{20}$$

$$s_{fix} = -\frac{3\eta}{\gamma_0} \frac{1}{\sqrt{\bar{\gamma}_0 Y}} \quad , \quad k_{fix} = \frac{3(4\eta^2 - \gamma_0^2)}{\gamma_0^2 \bar{\gamma}_0} \frac{1}{Y} \tag{21}$$

With $\eta = 0$, we obtain back the DLA results.

## 3. Relating parton and hadron spectra

In the theoretical calculations, we take the partons as massless, but introduce a $p_t$ cutoff $Q_0$ for regularization; on the other hand, the observable hadrons are massive. One can make the assumption that the cutoff $Q_0$ is related to the masses of hadrons. As a first stage in the discussion of charged particle spectra, one can relate $Q_0$ to an effective hadron mass[12], then for both partons and hadrons $E \geq Q_0$. Furthermore we relate the spectra

$$E_h \frac{dn(\xi_E)}{dp_h} = E_p \frac{dn(\xi_E)}{dp_p} \quad ; \quad E_h = E_p \geq Q_0 \tag{22}$$

at the same energy $E$ or $\xi_E = \log P/E$ (not momentum!), where $E_h = \sqrt{p_h^2 + Q_0^2}$ and $E_p = p_p$. With this choice both spectra vanish linearly at the same $E$ or $\xi_E$. To see this, we note that the hadronic data behave approximately like

$$E \frac{dn}{d^3p} \sim e^{-E/E_o} \quad \text{or} \quad E \frac{dn}{dp} \simeq 8\pi Q_0 (E - Q_0) e^{-E/E_0} \tag{23}$$

for sufficiently small energies (see Figure 3a below). The same linear decrease with energy towards the kinematic limit is also followed by the parton distribution, $Edn/dp \sim \log E/Q_0 \sim (E - Q_0)/Q_0$ (eq. (14)) and corresponds to the limited phase space with $p_p > Q_0$.

## 4. Results

*Moments.* The moments $<\xi^q>$ are determined from the spectra $Edn/dp$ vs. $\xi_E$ after appropriate transformation of the measured $x_p = 2p/\sqrt{s}$ spectra of charged particles using $E = \sqrt{p^2 + Q_0^2}$ and therefore depend on $Q_0$. For the unmeasured interval near $\xi_E \simeq Y$ (small momenta) a contribution was found by extrapolation. The error for the moments includes the errors on the central values of $\xi_E$ in each bin, taken as half the bin-size. In the same way we also obtain the multiplicity $\mathcal{N}_E$ as integral over $\xi_E$ of the full spectrum $Edn/dp$; its difference to the usual particle multiplicity $\bar{n}$ depends on the c.m. energy, ranging from a decrease of 30% at $\sqrt{s} = 3$ GeV to a decrease of 10% at LEP energy[5]. First we perform a fit of the two parameters $Q_0$ and $\Lambda$ by comparing the moments determined for a selected $Q_0$ with the theoretical predictions from eq. (9) of different $\Lambda$. The best agreement is obtained for

$$Q_0 \simeq \Lambda \simeq 270 \text{ MeV} \tag{24}$$

---

[5] the MARK I data point at $\sqrt{s} = 4.03$ GeV shows an anomalous decrease up to 50%, which may be partly related to charm thresholds effects.

and we estimate $\Delta Q_0 \simeq \Delta \Lambda \simeq 20$ MeV. The cumulants up to order 4 together with the corresponding MLLA predictions are shown in Figure 1. We also analyzed in the same way results from ADONE[17] at $\sqrt{s} = 1.6$ GeV which however reports only charged pions and therefore cannot necessarily be considered on the same ground as the other data. Nevertheless it is interesting to observe that the results follow smoothly the trend of the higher energy data, in particular $\bar{\xi}_E = 0.53 \pm 0.07$ and $\sigma^2 = 0.07 \pm 0.02$. Errors of the cumulants are taken as the sum of statistical and systematic errors. The latter particularly affects the overall normalization and then the average multiplicity; higher order cumulants are on the contrary independent of the overall normalization. The LEP average has been obtained by following the PDG procedure[18] for weighted averages. The MLLA predictions are calculated for the number of flavors $n_f = 3$. Including the heavy flavors at scale $P/4$ as discussed above would lower the prediction of $\sigma^2$ at LEP by two std. and by less than one std. at lower energies; likewise the effect on $\bar{\xi}_E$, $s$, and $k$ is negligible.

The result $Q_0 \simeq \Lambda$ ("limiting spectrum") follows, as an increase of $\lambda = \log(Q_0/\Lambda)$ would predict lower values for $\bar{\xi}_E$ but larger values for $\sigma^2$. These changes cannot be cured by a lowering of $Q_0$, which would shift both $\bar{\xi}_E$ and $\sigma^2$ downwards. This can be seen from the first two terms in the expansion of $\bar{\xi}_E$ and $\sigma^2$: $\bar{\xi} = Y/2 + ab(\sqrt{Y+\lambda} - \sqrt{\lambda})/24$ and $\sigma^2 = ((Y+\lambda)^{3/2} - \lambda^{3/2})/6b - bY/96$. For the multiplicity we write $\bar{\mathcal{N}}_E = c_1 \frac{4}{9} 2 \bar{\mathcal{N}}_{LS} + c_2$ with arbitrary normalization parameters for $\bar{\mathcal{N}}_{LS}$ in eq. (11) (the factor 4/9 is for the quark jet and 2 is for the two hemispheres) and the leading particle contribution. Fitting the lowest and highest energy data points we find $c_1 = 1.43$ and $c_2 = 1.02$.

We have not been able to obtain a satisfying description for the fixed-$\alpha_s$ case. We show in Fig. 1 one illustrative example for the same $Q_0 = 270$ MeV and $\gamma_0 = 0.64$ (with $n_f = 3$). With this choice for $\gamma_0$ one obtains a good fit for the multiplicity, where we proceed as above and we write $\bar{\mathcal{N}}_E = c_1 \frac{4}{9} 2 \bar{\mathcal{N}}_{fix} + c_2$ using eq. (17). In this case $c_1 = 3.10$ and $c_2 = -1.24$. Also the asymptotic slope for $\bar{\xi}_E$ is well reproduced. An adjustement of the absolute normalization of the moments $\bar{\xi}_E$ at a particular energy $Y_0$ is possible if the PQCD evolution towards lower energies and the normalization at threshold are abandoned.

The predictions from MLLA with running $\alpha_s$ and normalization at threshold are remarkably successful considering the fact that there are only two parameters, actually coinciding, for the four moments. A deviation is visible for $\bar{\xi}_E$ at the lower energies; this may indicate the influence of the leading valence quark which is neglected by restricting to the $D^+$ spectrum. Otherwise the predictions are confirmed at an almost quantitative level down to the lowest energy where charged particle spectra are measured. Also the $\log \bar{\mathcal{N}}_E$ data show a clear downward bending predicted from the running $\alpha_s$ if the lower energy data points are included. The distinction from the fixed $\alpha_s$ case would however require still higher energies. Excluding the lowest two data points of $\bar{\xi}$, we find an overall $\chi^2/NDF \simeq 1.8$.

On the other hand, the same model but with fixed coupling only fits the data on $\bar{\mathcal{N}}_E$ and the slope of $\bar{\xi}_E$ but fails otherwise. A fixed $\alpha_s$ regime can be excluded already close to threshold. Indeed, if we suppose that the coupling is fixed for a certain energy interval near threshold $0 \leq Y \leq Y_1$ and only runs for $Y > Y_1$, then $\bar{\xi}$ would be shifted towards smaller values because of the very different evolution near threshold; for example for $Y_1 = 0.3$ ($\sqrt{s} \simeq 0.7$ GeV) the shift would be $\Delta \bar{\xi} \simeq 0.25$, considerably away from the data. Also, if we restrict the discussion to higher energies, one cannot fit the higher moments with the same parameter $\gamma_0$ used for $\bar{\mathcal{N}}_E$ and $\bar{\xi}_E$, as can be better seen from the rescaled cumulants.

*Rescaled cumulants.* In addition to the standard moment analysis performed in the previous subsection, we also consider the rescaled cumulants $\kappa_q/\bar{\xi}$. These quantities become energy independent in case of fixed coupling (see eq. (8)) at high energies and therefore exhibit more directly the difference to the case of running coupling. In Figure 2 we show the experimental results on these ratios which have a considerable $Y$-dependence and are well reproduced by MLLA with running $\alpha_s$ in the available energy range. The Figure also shows the effect of including the heavy quarks at scale $P/4$ as discussed above. It would be interesting to continue this type of studies with data from jets of higher energies, where the MLLA predictions for these ratios continue to show a remarkable dependence on $Y$.

*Low-energy behavior: Boltzmann factor and running-$\alpha_s$ effect.* Let us consider the Lorentz-invariant distribution $Edn/d^3p$ as a function of the particle energy $E$ for small energies (up to 1 GeV). Experimental data from low energy experiments show an exponential decrease as in eq. (23) with $E_0$ of the order of 150 MeV[19, 17, 20]. This behavior is often related to the Boltzmann factor in the thermodynamical description of multiparticle production as, for example, in Hagedorn's model[21].

Figure 3a shows experimental data on the invariant distribution at three different c.m. energies. All three distributions tend to a common behavior at very low energies $E$, but deviations from a simple exponential are visible at larger c.m. energies. In Figure 3b, we also show the theoretical predictions for the MLLA with running and fixed $\alpha_s$ at $\sqrt{s}$ = 3 GeV. The running-$\alpha_s$ prediction has been obtained by using the distorted Gaussian parametrization as in [5] with the first four moments computed for the "limiting spectrum" with the same value of $Q_0$. The fixed-$\alpha_s$ curve is obtained from the exact solution for the energy spectrum given in eq. (14) again with $\gamma_0 = 0.64$. Both theoretical predictions are then normalized to the experimental average multiplicity.

It is remarkable that the MLLA with running $\alpha_s$ is in agreement with experimental data even at such a low energy; the exponential decrease is well reproduced by the running-$\alpha_s$ effect, which mimics a thermodynamical behavior. On the contrary, the fixed-$\alpha_s$ model shows a flatter behavior, inconsistent with experimental behavior. As for decreasing particle energy $E$ also the typical particle $p_T$ is necessarily decreasing, the coupling $\alpha_s(p_T/\Lambda)$ is rising in the running $\alpha_s$ case. This yields the steeper slope in comparison to the fixed $\alpha_s$ case. Therefore the running of the QCD coupling is visible even in the energy range of few hundreds MeV in the particle spectra.

The behavior near the kinematical limit $E \to Q_0$ can be easily understood analytically in the DLA. The first iteration of the evolution equation of the energy spectra[2] (the Born term) yields a finite contribution for $E \to Q_0$ and does not depend on c.m. energy whereas the second term provides the rise of the spectrum for large $E$ with increasing $\sqrt{s}$ for both fixed and running $\alpha_s$[15] in qualitative agreement with Figure 3a. For the ratios of invariant cross sections, $I(E) \equiv Edn/d^3p$, we find for small $E$:

$$\frac{I_{run}(E)}{I_{fix}(E)} = \frac{12}{\lambda b \gamma_0^2}\left(1 - \frac{1}{2}\frac{\log E/Q_0}{\lambda} + \ldots\right) \qquad (25)$$

There is a large relative enhancement of this ratio from the running of $\alpha_s$ at small $E \gtrsim Q_0$ followed by a steep decrease for typical parameters ($12/b\gamma_0^2 \simeq 4.3$ for $\gamma_0 = 0.64$, $\lambda < 1$), as the $p_t$ integral of the Born term is dominated by small values of $p_t < E$ and therefore large coupling $\alpha_s(p_t)$.

## 5. Conclusions and final remarks

The dependence on c.m. energy of the experimental first four order moments of the inclusive log $1/x$ distribution is well reproduced by the MLLA of perturbative QCD applied to the full c.m. energy region down to threshold. We found that the best agreement with data is for the limiting spectrum, where $Q_0 = \Lambda$; the best value of $Q_0$ has been estimated to be around 270 MeV and corresponds to an effective mass of a charged particle. The zero-th order moment, i.e., the average multiplicity, requires two more parameters which fix the overall normalization for particle production and the leading particle contribution. The effect of the running $\alpha_s$ is most pronounced near threshold where the variation of the coupling is strongest; it is also directly seen in the steepness of the slope of the invariant energy spectrum at low particle energies $E \lesssim 1$ GeV. Predictions of the same model but with fixed coupling are inconsistent with the experimental behavior in the full energy range. We take the evidence for the running $\alpha_s$ effects at low energies $E$ and $\sqrt{s}$ as further support for the local parton hadron duality hypothesis.

The distribution of particles at low energies $\sqrt{s} \lesssim 3$ GeV is not just according to phase space: the running $\alpha_s$ enhances configurations with collimated partons. At the hadronic level such configurations correspond to hadronic resonances which are known to dominate the final states already at these low energies. It is therefore suggested to consider a duality between partonic and hadronic distributions already at low energies. The running $\alpha_s$ with strong coupling at small $p_T$ seems to be related to the production of resonances.

A word of caution is appropriate nevertheless. The MLLA results used here are based on certain high energy approximations which might not be fully justified at low energies (restriction to $D^+$ contribution, neglect of Next-to-Next-to-Leading effects). A deviation in $\bar{\xi}$ at 3 GeV could hint towards a need of further contributions. Also two loop effects, whose importance increases towards lower energies, have been neglected in the calculation. At the price of introducing new parameters one could of course restrict the comparison with data to an energy range $Y \geq Y_0 > 0$ above threshold. At any rate, the high energy approximations work well for energies $\gtrsim 3$ GeV, where data on charged particle distributions have been presented.

The higher moments $q \geq 2$ show a remarkable energy dependence in the energy range above the LEP experiments. This suggests that the analysis of hadronic jets along these lines at the TEVATRON at ten times higher energies or at future colliders should be able to test the proper running of $\alpha_s$ towards much higher energies.

## 6. Acknowledgements

Helpful discussions with V. A. Khoze, S. Pokorski and D. V. Shirkov are gratefully acknowledged. One of us (S.L.) would like to thank D.A.A.D. for financial support.

**Figure Captions**

**Fig. 1**: The average multiplicity $\bar{\mathcal{N}}_E$ and the first four cumulants of charged particles' energy spectra $E dn/dp$ vs. $\xi_E$, i.e., the average value $\bar{\xi}_E$, the dispersion $\sigma^2$, the skewness $s$ and the kurtosis $k$, are shown as a function of $Y = \log(\sqrt{s}/2Q_0)$ for $Q_0 = 270$ MeV. Data points from MARK I[20] at $\sqrt{s} = 3$, 4.03, 7.4 GeV, TASSO[7] at $\sqrt{s} = 14$, 22, 35, 44 GeV and LEP at $\sqrt{s} = 91.2$ GeV (weighted averages of ALEPH[10], DELPHI[11], L3[9] and OPAL[8]). Predictions of the "limiting spectrum" (i.e. $Q_0 = \Lambda$) of MLLA with running $\alpha_s$ and of MLLA with fixed $\alpha_s$ are also shown (for $n_f = 3$). Predictions of the average multiplicity refer to the two-parameter formula $\bar{\mathcal{N}}_E = c_1 \frac{42}{9} \bar{\mathcal{N}} + c_2$.

**Fig. 2**: Rescaled cumulants $\kappa_q/\bar{\xi}$ as a function of $Y = \log(\sqrt{s}/2Q_0)$ with the corresponding predictions of the "limiting spectrum" of MLLA with running $\alpha_s$ either without or with heavy flavors included. Data as in Figure 1. For fixed $\alpha_s$ these quantities approach constant values at high energies.

**Fig. 3**: Invariant distribution $E dn/d^3p$ as a function of the particle energy $E$ for $Q_0 = 270$ MeV. **a)**: experimental results from MARK I[20], TASSO[7] and OPAL[8]; **b)**: comparison of experimental data from MARK I[20] with predictions from MLLA with running $\alpha_s$ and with fixed $\alpha_s$ at the same energy.

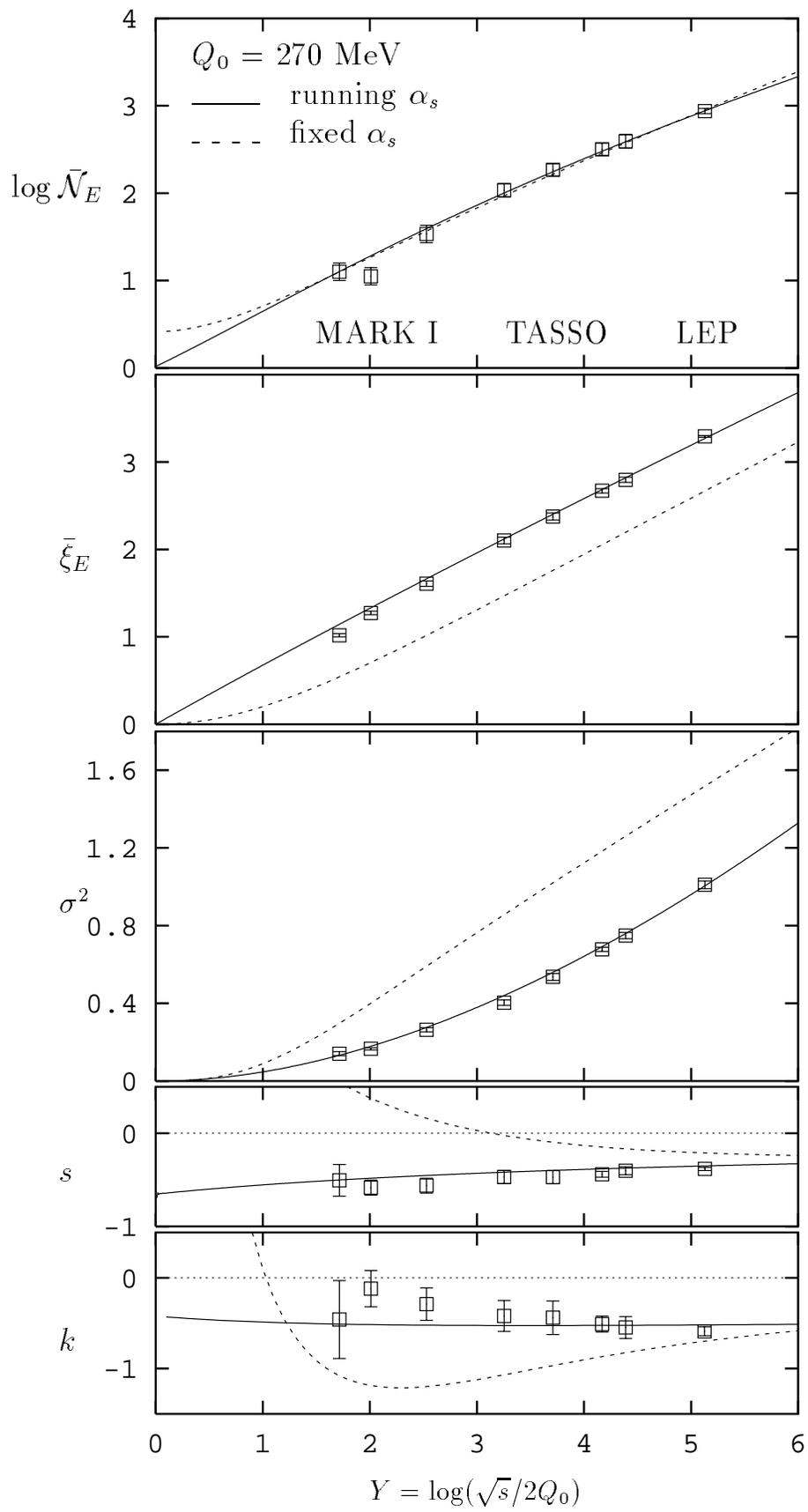

Figure 1

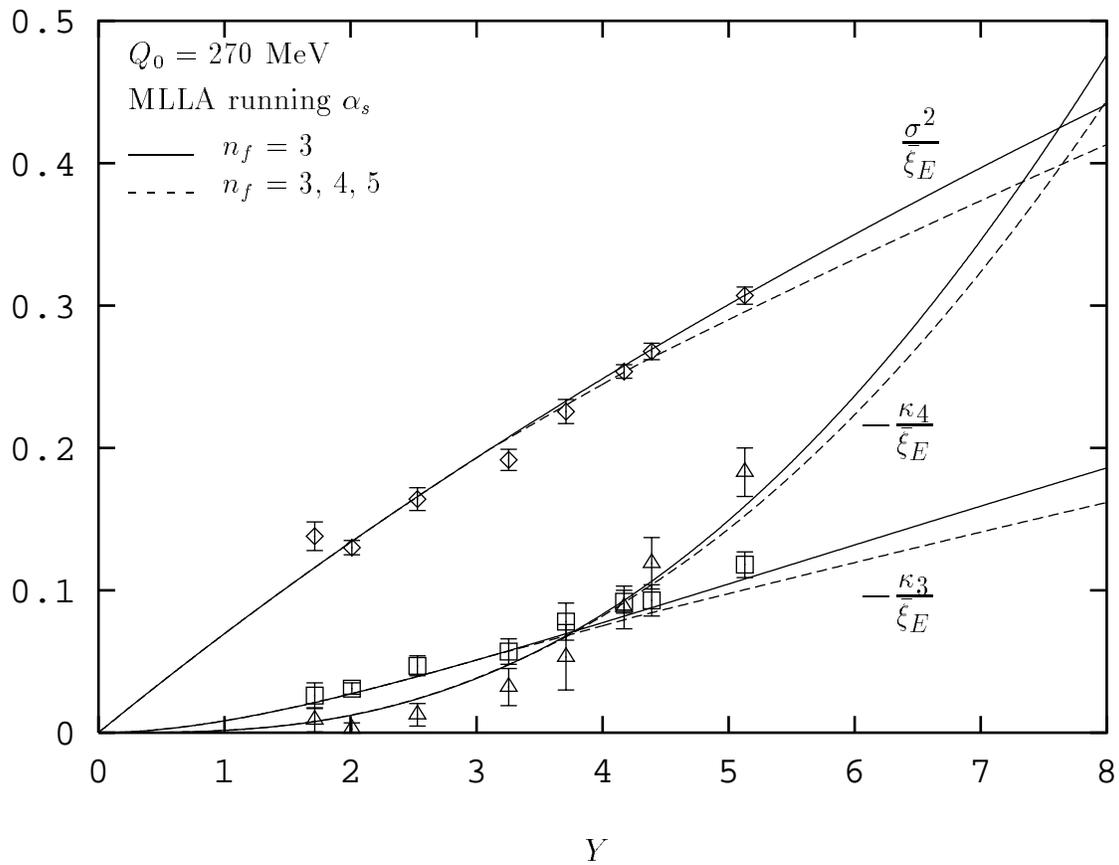

Figure 2

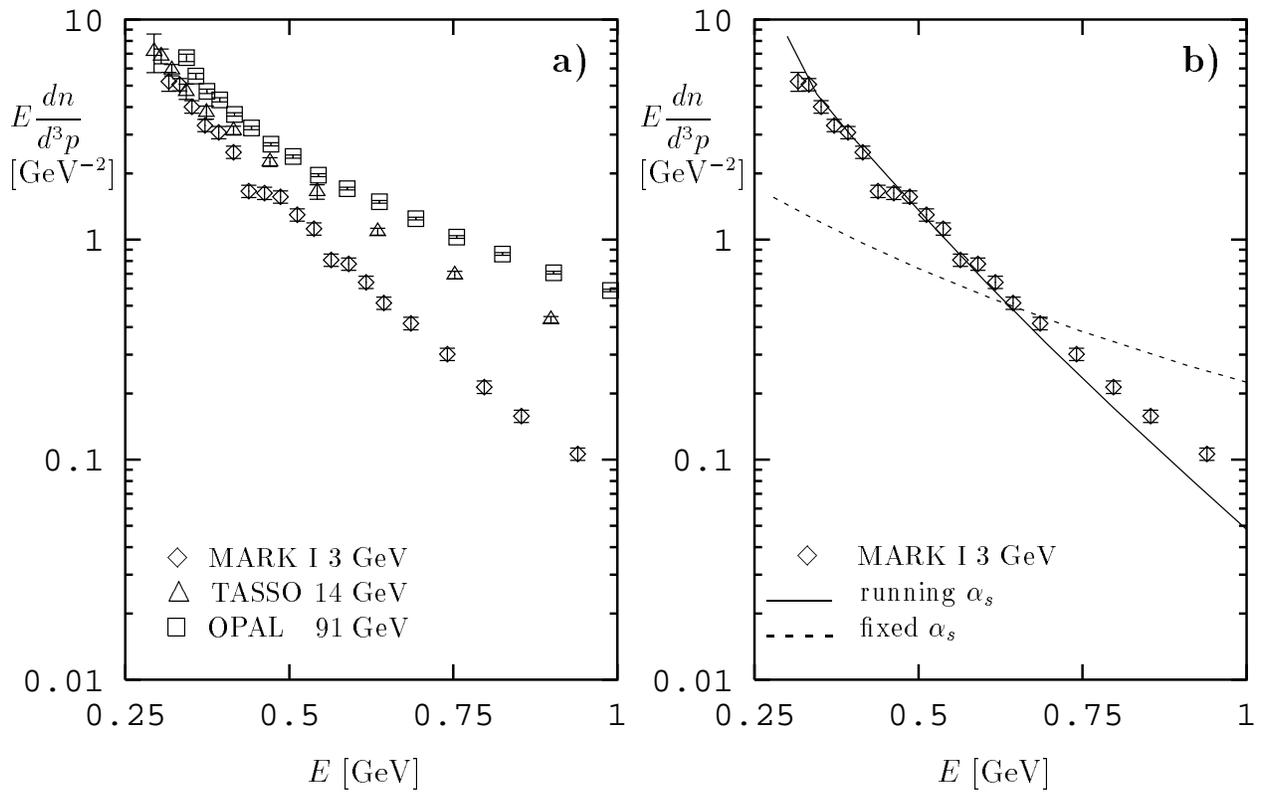

Figure 3